\documentclass[fleqn,usenatbib]{mnras}

\usepackage{newtxtext,newtxmath}
\usepackage[T1]{fontenc}
\usepackage{ae,aecompl}

\usepackage{float}
\usepackage[caption = false]{subfig}
\usepackage{graphicx}	% Including figure files
\usepackage{amsmath}	% Advanced maths commands
\usepackage{amssymb}	% Extra maths symbols
\usepackage{booktabs}
\usepackage{color}

\title[Cosmic voids uncovered]{Cosmic voids uncovered -- first-order statistics of depressions in the biased density field}

\author[T. Ronconi et al.]{
T. Ronconi,$^{1,2}$\thanks{E-mail: tronconi@sissa.it}
S. Contarini,$^{3,4}$
F. Marulli,$^{3,4,5}$
M. Baldi$^{3,4,5}$
and L. Moscardini$^{3,4,5}$
\\
$^{1}$SISSA - International School for Advanced Studies, via Bonomea 265, I-34136 Trieste, Italy\\
$^{2}$INFN - Sezione di Trieste, via Valerio 2, I-34127 Trieste, Italy\\
$^{3}$Dipartimento di Fisica e Astronomia - Alma Mater Studiorum Universit\`{a} di Bologna, via Piero Gobetti 93/2, I-40129 Bologna, Italy\\
$^{4}$INAF - Osservatorio di Astrofisica e Scienza dello Spazio di Bologna, via Piero Gobetti 93/3, I-40129 Bologna, Italy\\
$^{5}$INFN- Sezione di Bologna, viale Berti Pichat 6/2, I-40127 Bologna, Italy
}

\date{Accepted XXX. Received YYY; in original form ZZZ}

\pubyear{2018}

\begin{document}
\label{firstpage}
\pagerange{\pageref{firstpage}--\pageref{lastpage}}
\maketitle

\begin{abstract}
Cosmic voids are the major volume component in the matter distribution of the Universe.
They posses great potential for constraining dark energy as well as for testing theories of gravity.
Nevertheless, in spite of their growing popularity as cosmological probes, a gap of knowledge between cosmic void observations and theory still persists.
In particular, the void size function models proposed in literature have been proven unsuccessful in reproducing the results obtained from cosmological simulations in which cosmic voids are detected from biased tracers of the density field, undermining the possibility of using them as cosmological probes.
The goal of this work is to cover this gap. In particular, we make use of the findings of a previous work in which we have improved the void selection procedure, presenting an algorithm that redefines the void ridges and, consequently, their radius.
By applying this algorithm, we validate the volume conserving model of the void size function on a set of unbiased simulated density field tracers.
We highlight the difference in the internal structure between voids selected in this way and those identified by the popular {\small VIDE} void finder.
We also extend the validation of the model to the case of biased tracers.
We find that a relation exists between the tracer used to sample the underlying dark matter density field and its unbiased counterpart.
Moreover, we demonstrate that, as long as this relation is accounted for, the size function is a viable approach for studying cosmology with cosmic voids.
\end{abstract}

\begin{keywords}
large-scale structure -- cosmology:theory -- methods:statistical
\end{keywords}

%%%%%%%%%%%%%%%%%%%%%%%%%%%%%%%%%%%%%%%%%%%%%%%%%%

%%%%%%%%%%%%%%%%% BODY OF PAPER %%%%%%%%%%%%%%%%%%

\section{Introduction}

Cosmic voids are large underdense regions where matter is evacuated and then squeezed in between their boundaries.
 While galaxy clusters encapsulate most of the mass in the Universe, voids are the dominant component of the Universe volume.
 Voids are only mildly non-linear objects since, during their evolution, they encounter a crucial physical limit: matter density cannot be less than zero, in other words, the minimum possible density contrast achievable by a region is $\delta = -1$.
Conversely to what happens to overdensities, the void evolution tends to reduce the possible non-sphericity of the initial density perturbations \citep{Icke1984}.
These properties suggest that the employment of a simplified spherical expansion model may be more accurate in describing the void formation and evolution than it is in the description of overdensities \citep{Blumenthal1992}.

In the last decade, cosmic voids have gathered a large popularity as cosmological tools.
With typical sizes over tens of megaparsecs, they are by far the largest observable structures in the Universe.
For this reason they are particularly suited to provide information about several hot topics.
They are indeed extreme objects, meaning that they generate from the longest wavelengths of the matter perturbation power spectrum.
Moreover, their underdense nature makes them suitable for a wide range of dark energy measurements.
Indeed, the {\em super-Hubble velocity field} within voids causes a suppression in structure growth, making them a pristine environment in which to study structure formation \citep{Benson2003, VandeWeygaert2011}, and a test-bench for constraining dark matter (DM), dark energy and theories of gravity \citep{Lavaux2012, Li2012, Sutter2014, Massara2015, Pollina2016, Kreisch2018}.

The excursion-set model has been applied to the study of underdensities, in particular to predict the size function of cosmic voids detected in the DM field \citep[][hereafter SvdW]{SVdW2004}.
It is based on the assumption that cosmic voids are nearly spherical regions which have gone through shell-crossing. 
The latter represents the time when the inner, more underdense, matter shells inside voids have reached the outer, as a result of the differential outward acceleration they experience. \cite{Jennings2013} then proposed a modification of this model that makes it in better agreement with the actual size distribution of cosmic voids measured in both simulated and real void catalogues. 

The first goal of our work is to further validate the \citet{Jennings2013} size function model on simulated void catalogues detected from unbiased density field tracers, at different redshifts and numerical resolutions. Then we extend the model to account for the more realistic case of biased tracers, covering the existing gap between the theoretical and observational sides of the problem.

Our work is organised as follows.
In Section \ref{sec:method} the existing void size function theoretical models are presented.
We focus in particular on their realm of applicability, underlining the existing issues in modelling void statistics in biased matter distributions. 
We then propose a modification of the \citet{Jennings2013} void size function model to account for the tracer bias.
Section \ref{sec:results} is dedicated to the validation of our fiducial model. We investigate its reliability using void catalogues extracted from a set of cosmological N-body simulations with different selections.
Specifically, while in Section \ref{sec:unbiasedtracers} we make use of DM particle simulations, meaning that voids are detected from unbiased tracers of the density field, in Section \ref{sec:biasedtracers} we extend our study to voids identified in the distribution of biased DM haloes.
Finally, in Section \ref{sec:conclusions} we draw our conclusions.

%%%%%%%%%%%%%%%%%%%%%%%%%%%%%%%%%%%%%%%%%%%%%%%%%%%%%%%%%%%%%%%%%%%%%%%%%%%%%%%
%%%%%%%%%%%%%%%%%%%%%%%%%%%%%%%%%%%%%%%%%%%%%%%%%%%%%%%%%%%%%%%%%%%%%%%%%%%%%%%
%%%%%%%%%%%%%%%%%%%%%%%%%%%%%%%%%%%%%%%%%%%%%%%%%%%%%%%%%%%%%%%%%%%%%%%%%%%%%%%

\section{Methodology}
\label{sec:method}

Whether it is possible to exploit cosmic voids as cosmological probes is a matter of how reliable is our capability to model their properties.
The most straightforward measure we can compute from whatever survey of extragalactic sources is the abundance of sources as a function of a particular feature. In the case of cosmic voids, this is the size function.

In Section \ref{sec:vsf}, we first present the existing theoretical models for the size function of cosmic voids extracted from unbiased tracers of the DM density field. Then we introduce our new parameterisation to describe the void size function in the more realistic case of void catalogues extracted from biased tracer distributions. In Section \ref{sec:finder}, we introduce the detailed procedure that we will use later on to identify the cosmic voids in a set of DM-only N-body simulations, and to clean the catalogue.

The simulations employed in the present work have been performed with the {\small{C-GADGET} module} \citep[][]{Baldi_etal_2010}, a modified version of the (non-public) {\small GADGET3} N-body code \citep[][]{gadget}, while the  software exploited for the post-processing and the data analysis\footnote{Specifically, we use the numerical tools to handle cosmic void catalogues provided by the {\small CosmoBolognaLib}, V5.1 \citep{Marulli2016}. This consists of a large set of C++/python {\em free software} libraries, freely available at the following GitHub repository: \url{https://github.com/federicomarulli/CosmoBolognaLib}.} has been presented in \citet[hereafter RM17]{RonconiMarulli2017}, where we also provided a first description of our new theoretical size function model and cleaning method.

%%%%%%%%%%%%%%%%%%%%%%%%%%%%%%%%%%%%%%%%%%%%%%%%%%%%%%%%%%%%%%%%%%%%%%%%%%%%%%%
%%%%%%%%%%%%%%%%%%%%%%%%%%%%%%%%%%%%%%%%%%%%%%%%%%%%%%%%%%%%%%%%%%%%%%%%%%%%%%%

\subsection {Size function}
\label {sec:vsf}

The development of any theoretical model of void size function relies on the definition of cosmic void.
Let us stick with the common assumption that an isolated void evolves from an initial spherical top-hat depression in the density field.
This is analogous to the assumption that an isolated DM halo results from the spherical collapse of a top-hat peak in the density field.
The evolution of both overdensities (peaks) and underdensities (depressions) can be described via the classical spherical evolution model \citep{Blumenthal1992}.
In the overdensity case, a halo is said to have formed at the moment its density contrast reaches a level corresponding to either the virialization of the spherical perturbation \cite[the well-known critical linear overdensity $\delta_c \approx 1.69$, for an EdS Universe, see e.g.][]{Paranjape2012} or, with a milder assumption, when the perturbation reaches the {\em turn-around} ($\delta_t \approx 1.06$, for an EdS Universe\footnote{We will generically refer to both the {\em critical} and the {\em turn-around} densities with the same symbol, $\delta_c$.}), and detaches from the overall expansion of the Universe.

On the other hand, underdensities do never detach from the overall expansion, but instead they expand with a {\em super-Hubble flow}.
The expansion rate is inversely proportional to the embedded density, therefore shells centred around the same point are expected to expand faster as more as they are close to the centre.
This eventually leads to the inner shells reaching the outer, the so-called {\em shell-crossing}, which is typically assumed to be the moment when a void can be said to have formed.
In linear theory, the shell-crossing occurs at a fixed level of density contrast ($\delta_v^L\approx -2.71$, for an EdS Universe\footnote{We will use the superscript L to identify linearly extrapolated quantities, while the superscript NL will mark their non-linear counterparts.}). This threshold can be used in an excursion-set framework to predict the typical distribution of voids, analogously to the case of halo formation by spherical collapse \citep{Bond1991,Zentner2007}.
With these assumptions, SvdW developed a model of the void size function that, furthermore, considers the {\em void-in-cloud} side effect, which accounts for the squeezing of voids that happen to evolve within larger scale overdensities.

In this framework, the probability distribution of voids of a certain size is given by
\begin{equation}
  \label{eq:vsf01}
  f_{\ln\sigma} = 2 \sum_{j=1}^{\infty}j \pi x^2 \sin(j \pi
  \Delta)\exp\biggl[-\frac{(j \pi x)^2}{2}\biggr]\, ,
\end{equation}
where
\begin{equation}
  x = \frac{\Delta}{|\delta_v^L|}\,\sigma 
\end{equation}
and
\begin{equation}
  \Delta = \frac{|\delta_v^L|}{\delta_c + |\delta_v^L|}\, .
\end{equation}
In the previous equation $\sigma$ is the square root of the variance, computed in terms of the size of the considered region:
\begin{equation}
\sigma^2=\frac{1}{2\pi}\int k^2P(k) W^2(k,r)\text{d} k \, ,
\end{equation}
where $P(k)$ is the matter power spectrum, $W(k,r)$ the window function, and $r$ is the radius of the spherical underdense region defined as void.

With the kernel probability distribution of Eq. \ref{eq:vsf01}, it is straightforward, in linear theory, to obtain the number density distribution of voids as a function of their size by applying:
\begin{equation}
  \label{eq:vsf02}
  \frac{\text{d}n}{\text{d}\ln r}\biggr|_{linear} = \frac{f_{\ln\sigma} (\sigma)}{V(r)} \frac{\text{d}\,\ln\sigma^{-1}}{\text{d}\,\ln r}\ \text{.}
\end{equation}
In order to derive the void size function in the non-linear regime, a conservation criterion has to be applied.
The SvdW size function relies on the assumption that the total number of voids is conserved when going from linear to non-linear\footnote{This is the same assumption which is implicitly made to derive the halo mass function.}.
While reaching shell-crossing, underdensities are expected to have expanded by a factor $a \propto (1 + \delta_v^{NL})^{-1/3}$, thus a correction in radius by this factor is required:
\begin{equation}
  \label{eq:vsf03}
    \frac{\text{d}\,n}{\text{d}\, \ln r} \biggr|_{\text{SvdW}}
    = \frac{\text{d}\,n}{\text{d}\, \ln (a\,r)}\biggr|_{linear}\ \text{.}  
\end{equation}
\cite{Jennings2013} argued that such a prescription is unphysical, since this leads to a volume fraction occupied by voids which is larger than the total volume of the Universe.
They thus introduced a {\em volume conserving} model (hereafter Vdn model) in which it is instead assumed that the total volume occupied by cosmic voids is conserved in the transition to non-linearity.
Nonetheless, when shell-crossing is reached, voids are thought to recover the overall expansion rate, and continue growing with the Hubble flow \citep{Blumenthal1992,SVdW2004}.
The conservation of volume is achieved by appling:
\begin{equation}
  \label{eq:vsf04}
  \frac{\text{d}\,n}{\text{d}\, \ln r}\biggr|_{\text{Vdn}} =
  \frac{\text{d}\,n}{\text{d}\, \ln r}\biggr|_{\text{linear}} \frac{V(r^L)}{V(r)}
  \frac{\text{d}\, \ln r^L}{\text{d}\, \ln r}\ \text{,}
\end{equation}
where $r^L$ indicates the radius predicted by linear theory (i.e. not accounting for the conversion factor $a$).

Several authors have tested the SvdW model on both simulated DM density fields and in mock halo catalogues, finding out that it systematically underpredicts the void comoving number density \citep[see e.g.][]{Colberg2005,Sutter2012,Pisani2015,Nadathur2015}.
To overcome this mismatch, the underdensity threshold $\delta_v^L$ is commonly left as a free parameter, tuned on simulated halo catalogues.
This severely affects the possibility of using the void size function as a cosmological probe.
\cite{Jennings2013} have shown that the Vdn model does not require such a fine-tuning, as long as the void catalogue is properly cleaned from spurious voids.
However, their results are limited to the case of cosmic voids detected from simulated DM distributions. We extend their study to the case of biased samples, such as mock DM halo catalogues, which are more representative of the real case.

Several authors have underlined that the tracer bias has to be taken into account in order to extract unbiased cosmological information from the number counts of cosmic voids detected in galaxy redshift surveys \citep[see e.g.][]{Pollina2018}.
In fact, the comoving number density and sizes of voids traced by the DM distribution are different from the ones of the voids traced by a biased DM halo density field.

In RM17 we introduced a simple modification of the void size function model that accounts for this effect.
It follows from the results by \citet{Pollina2017}, that showed that the DM density field within voids, $\delta_{v,\, \text{DM}}^{NL}$, is linearly related to the density field traced by some biased tracers, $\delta_{v,\, \text{tr}}^{NL}$:
\begin{equation}
  \label{eq:bias01}
  \delta_{v,\, \text{tr}}^{NL} = b\, \delta_{v,\, \text{DM}}^{NL}\ \text{.}
\end{equation}
Therefore, the threshold density to be used in the size function model of cosmic voids detected from biased tracers has to be corrected taking into account Eq. \eqref{eq:bias01}.
Moreover, to recover the linearly extrapolated value $\delta_{v,\,\text{tr}}^{L}$, we use the fitting formula provided by \citet{Bernardeau1994}, namely 
\begin{equation}
\label{eq:bias03}
\delta^{NL} = (1 - \delta^L/\mathcal{C})^{-\mathcal{C}}\, ,
\end{equation}
with $\mathcal{C}=1.594$, which is valid for any cosmological model and is independent of the shape of the power spectrum with errors below $0.2\%$.
By combining it with Eq. \eqref{eq:bias01}, it gives
\begin{equation}
  \label{eq:bias02}
  \delta_{v,\,\text{tr}}^L = \mathcal{C}\, \bigl[1 - (1 + b\,
  \delta_{v,\, \text{DM}}^{NL})^{-1/\mathcal{C}} \bigr]\, .
\end{equation}
Our recipe for a void size function that works regardless of whether the void catalogue is affected or not by a tracer bias is to use the value of $\delta_v^L$ found with Eq. \eqref{eq:bias02} in the probability function given by Eq. \eqref{eq:vsf01}.

%%%%%%%%%%%%%%%%%%%%%%%%%%%%%%%%%%%%%%%%%%%%%%%%%%%%%%%%%%%%%%%%%%%%%%%%%%%%%%%
%%%%%%%%%%%%%%%%%%%%%%%%%%%%%%%%%%%%%%%%%%%%%%%%%%%%%%%%%%%%%%%%%%%%%%%%%%%%%%%

\subsection{Void finding and data reduction}
\label {sec:finder}

There is not general concordance in the definition of voids.
Indeed, many different techniques have been proposed and applied over the years \citep[see e.g.][]{Colberg2008, Elyiv2015}.
A significant part of our work has been dedicated to cover this gap of knowledge, through the development of a procedure to standardize the outcome of void finders of different types (see also RM17).
In this paper we use the public void finder {\small VIDE} \citep[][Void IDentification and Examination Toolkit]{SutterVIDE2015} to identify voids in our tracer distributions, whatever their nature and independently of the presence of a biasing factor.
The {\small VIDE} catalogues obtained are then {\em cleaned} with the pipeline developed and presented in RM17. This allows us to align the objects included in the void catalogue with the theoretical definition of cosmic voids used to derive the void size function (Sec. \ref{sec:vsf}).

We summarise here the main features of the cleaning pipeline, referring to RM17 for a detailed description.
The procedure is divided into three main steps: \textit{(i)} a preliminary trimming of the catalogue based on density and reliability criteria, \textit{(ii)} a rescaling of the effective radius of all the voids in the catalogue and, finally, \textit{(iii)} the catalogue is trimmed again to erase overlapping between voids.
As underlined also in RM17, this cleaning method is necessary to match the distribution of voids measured in simulated catalogues with the predictions of the Vdn model.

The rescaling part of the procedure (step \textit{(ii)}) is based on the requirement that cosmic voids have to embed a density matching the shell-crossing prescription defined by the model.
This is crucial in order to have a reliable description of their distribution in terms of spherical expansion.
In step \textit{(ii)} spheres are grown around the centre of voids selected by the adopted void finder algorithm up to the scale inside which an a-priori chosen value of underdensity is reached.
The latter value is typically chosen as the non-linear counterpart of the shell-crossing threshold, namely about $\delta_v^{NL} \approx -0.795$ at redshift $z = 0$ for an EdS Universe.

Even though it is reasonable to select this particular value as the critical threshold in both the model and the cleaning method, there are no restrictions in this sense and whatever value is acceptable, as long as it is used in both measuring (that is when applying the cleaning algorithm) and modelling the distribution (that is when building the modified model described in Sec. \ref{sec:vsf}).
This is a key issue: the choice of the underdensity threshold, that has to be embedded by the voids identified in whatever tracer distribution, must match the threshold $\delta_v^L$ in the model (Eq. \eqref{eq:vsf01}) used to predict their distribution.
Given that when measuring the size distribution of cosmic voids we are dealing with a non-linear Universe, while the theoretical model is derived from linear theory, a prescription to convert the threshold back and forth \cite[such as the fitting relation from][]{Bernardeau1994} is necessary.

%%%%%%%%%%%%%%%%%%%%%%%%%%%%%%%%%%%%%%%%%%%%%%%%%%%%%%%%%%%%%%%%%%%%%%%%%%%%%%%
%%%%%%%%%%%%%%%%%%%%%%%%%%%%%%%%%%%%%%%%%%%%%%%%%%%%%%%%%%%%%%%%%%%%%%%%%%%%%%%
%%%%%%%%%%%%%%%%%%%%%%%%%%%%%%%%%%%%%%%%%%%%%%%%%%%%%%%%%%%%%%%%%%%%%%%%%%%%%%%

\section{Results}
\label{sec:results}

We test the procedure outlined in Sec. \ref{sec:method} with both a set of DM-only N-body simulation snapshots, and a set of mock halo catalogues extracted from the same simulations by means of a Friends-of-Friends (FoF) algorithm, and applying different mass selection cuts.
Specifically, we consider a suite of N-body simulations of the standard $\Lambda$CDM cosmology performed with the {\small C-GADGET} module \citep[][]{Baldi_etal_2010} for different box sizes and particle numbers. Our largest box corresponds to the $\Lambda$CDM run of the {\small L-CoDECS} simulations \citep{baldi2012}, with a volume of $(1\, \text{Gpc}/h)^3$ and a total number of $2\cdot 1024^3$ particles, with a mass resolution of $\sim 6 \cdot 10^{10}\;M_\odot/h$. The smaller simulations have been run with the same code specifically for this work, and share with the largest box the same fiducial cosmological parameters \citep[consistent with WMAP7 constraints, see Table \ref{tab:cosmo_param}]{Komatsu_etal_2011}, spanning a range of lower, equal, and higher particle mass and spatial resolutions, with the aim of testing the dependence of our analysis on these numerical parameters.
\begin{table}
  \centering
  \caption{Fiducial cosmological parameters of the N-body simulations used in this work.}
  \begin{tabular}{cccccc}
    \toprule
    $h_0$ & $\Omega_{CDM}$ & $\Omega_b$ & $\Omega_{\Lambda}$ &     $\mathcal{A}_s$ & $n_s$    \\
    \midrule
    $0.7$  &       $0.226$ &   $0.0451$ &         $0.7289$ &  $2.194\cdot10^{-9}$ & $0.96$   \\
    \noalign{\smallskip}\hline
    \bottomrule
  \end{tabular}
  \label{tab:cosmo_param}
\end{table}
The main properties of the different simulations are reported in Table \ref{tab:simulations}.
\begin{table}
\centering
\caption{Our set of cosmological simulations with the corresponding relevant physical quantities: $L_{box}$ is the simulation box-side length in units of [Mpc$/h$]; $N_p$ is the total number of particles; m.i.s. is the mean inter-particle separation in the box in units of [Mpc$/h$]; in the last column, the particle mass in units of [$10^{10}\;M_\odot/h$] is reported. Note that, since it had been run with two matter components, the last listed simulation has two associated values which refer to the mass of the DM particles and of the baryonic particles, respectively.}
\begin{tabular}{lccr}
\toprule
\multicolumn{1}{c}{$L_{box}$} & \multicolumn{1}{c}{$N_p$} & \multicolumn{1}{c}{m.i.s.} & \multicolumn{1}{c}{particle mass} \\
\multicolumn{1}{c}{[Mpc$/h$]} & \multicolumn{1}{c}{} & \multicolumn{1}{c}{[Mpc$/h$]} & \multicolumn{1}{c}{[$10^{10}\;M_\odot/h$]} \\
\midrule
$500$ &        $256^3$ & $\approx 2.00$ &  $56.06$ \\
$256$ &        $256^3$ & $1.00$         &  $7.52$  \\
$128$ &        $256^3$ & $0.50$         &  $0.94$  \\
$64$ &        $256^3$ & $0.25$         &  $0.12$  \\
$1000$ & $2\cdot1024^3$ & $1.00$         &  $5.84, 1.17$  \\
\noalign{\smallskip}\hline
\bottomrule
\end{tabular}
\label{tab:simulations}
\end{table}

As a first step, in Section \ref{sec:unbiasedtracers} we validate our procedure on the four N-body simulations with $256^3$ particles listed in Table \ref{tab:simulations}. 
These simulations vary in mass resolution and mean inter-particle separation (m.i.s.).
We gathered snapshots at four different redshifts ($z = 0\text{,}\, 0.5\text{,}\, 1\text{,}\, 1.5$) for each of these four simulations.

Afterwards, in Section \ref{sec:biasedtracers} we extend our analysis to the case of biased tracers.
The set of halo catalogues is extracted from a snapshot at redshift $z = 0$ of the simulation with box-side length $L_{box} = 1000\ \text{Mpc}/h$.
We could not use the same simulations used in the validation part of the work because of the extremely low number of voids identified by {\small VIDE} in those small boxes.

%%%%%%%%%%%%%%%%%%%%%%%%%%%%%%%%%%%%%%%%%%%%%%%%%%%%%%%%%%%%%%%%%%%%%%%%%%%%%%%
%%%%%%%%%%%%%%%%%%%%%%%%%%%%%%%%%%%%%%%%%%%%%%%%%%%%%%%%%%%%%%%%%%%%%%%%%%%%%%%

\subsection {Unbiased tracers}
\label {sec:unbiasedtracers}

In order to validate our procedure, we first run the {\small VIDE} void finder on top of each of the 16 simulation snapshots described before with varying resolution and redshift. 
The resulting void catalogues are then cleaned and re-scaled with our pipeline.
We evolve the shell-crossing threshold, $\delta_v^L = -2.71$, as
\begin{equation}
    \label{eq:unbiased01}
    \delta_v^{L}(z) = \delta_v^{L}\cdot\dfrac{\mathcal{D}(z)}{\mathcal{D}(0)}\ \text{,}
\end{equation}
by means of the growth-factor $\mathcal{D}(z)$ at redshift $z$.
Applying the fitting formula given in Eq. \eqref{eq:bias03} we obtain its non-linear extrapolation, $\delta_v^{NL}(z)$, and use this value as a threshold in the cleaning pipeline.
Doing so, all the voids in our catalogues embed a fixed density 
\begin{equation*}
    \rho_v = \langle\rho\rangle \ \bigl[1 + \delta_v^{NL}(z)\bigr]\, ,
\end{equation*}
where $\langle\rho\rangle$ is the average matter density of the snapshot.

Firstly, we investigate the difference between the void density profiles before and after the application of our cleaning algorithm.
It is generally believed that, in the standard $\Lambda$CDM cosmology, voids have a self-similar internal structure, which does not depend on the tracers used to define them \citep[see e.g.][]{Ricciardelli2013, Ricciardelli2014, Hamaus2014}.
The self-similarity assures that the internal structure of voids can be characterised by a single parameter, the void effective radius, $r_\text{eff}$, which is the radius of a sphere embedding the same volume embedded by a given cosmic void.
Despite the wide range of values this parameter covers, it is possible to average the internal density distribution to recover a common behaviour for voids with approximately the same $r_\text{eff}$.

It has been noticed that, to recover self-similarity, some criteria have to be applied to the selection of objects to be classified as voids \citep{Nadathur2015}. 
Nevertheless, the most crucial role seems to be played by the cut on the minimum density of the candidates, $\rho_{min}$, which has to be sufficiently low (that is $\rho_\text{min} \lesssim 0.3\langle\rho\rangle$).
All voids undergoing our cleaning procedure with our chosen value of $\delta_v^{NL}$ satisfy this requirement by construction.
We are therefore confident that the objects we select should behave in a self-similar manner. 

We select all the cosmic voids in the range of radii $1\ \text{Mpc}/h < r_\text{eff} < 1.5\ \text{Mpc}/h$, from the catalogue with box side length $L_{box} = 128\ \text{Mpc}/h$, at four different redshifts. This is a compromise between having a large box (thus having more objects) and having a high resolution.
Nonetheless, the results we are going to present are similar for the other three simulations used in this section.
The radius window has been chosen to have a sufficiently high number of objects with radii far enough from the lower limit imposed by the spatial resolution of the box that, for this catalogue, is $0.5\ \text{Mpc}/h$ (see Table \ref{tab:simulations}).

\begin{figure*}
 \centering
\includegraphics[width=0.95\textwidth]{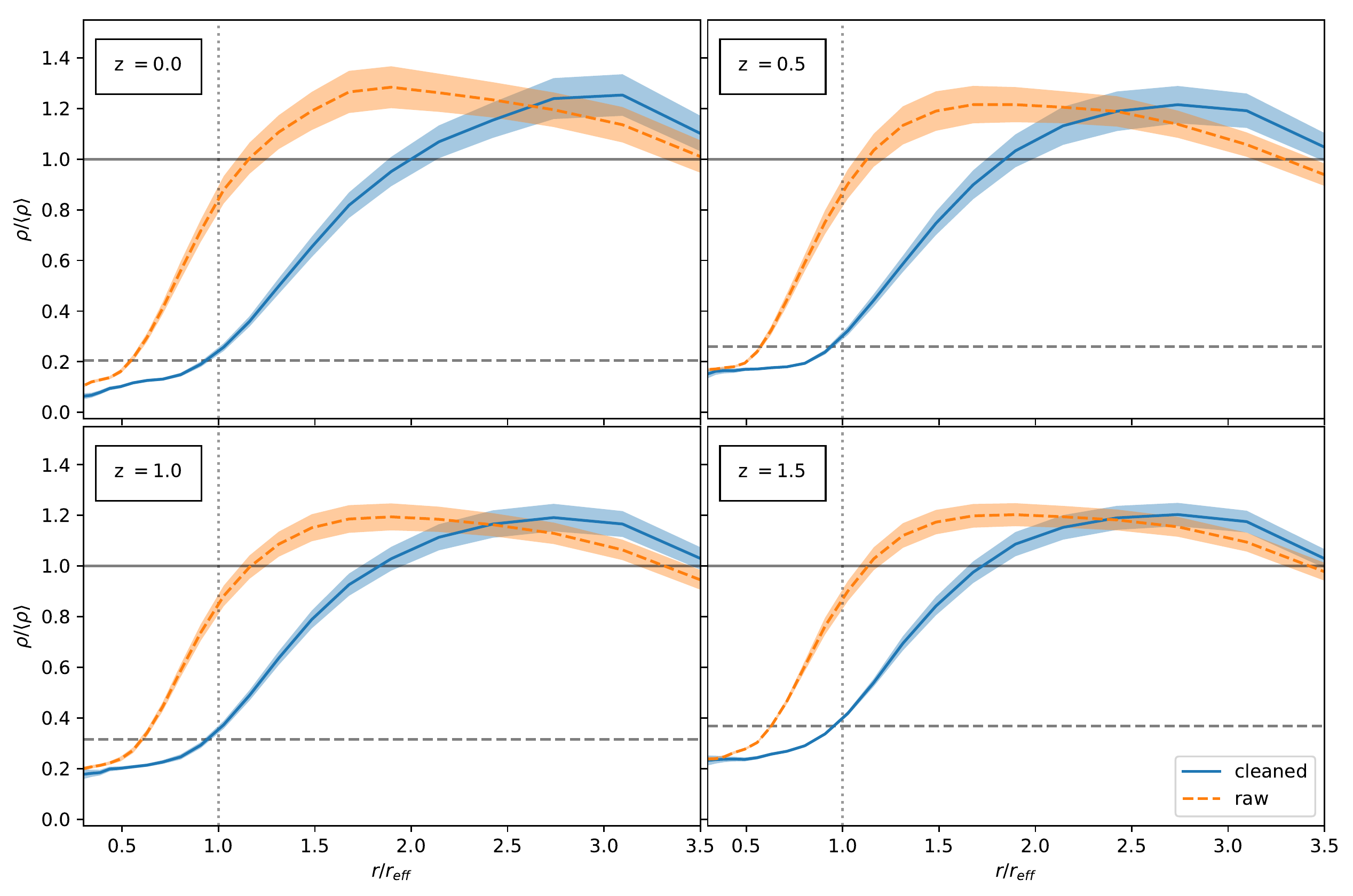}
\caption{Comparison between the density profile within voids obtained with {\small VIDE} before (dashed orange line) and after (solid blue line) applying our cleaning algorithm. Each panel shows the result of averaging all the density profiles in a narrow bin of radii (with $1 < r_\text{eff} [\text{Mpc}/h]< 1.5$) and the $2 \sigma$ confidence as a shaded region around the mean, at four different redshifts. The two horizontal lines show the position of the shell-crossing threshold (dashed line) and of the background density (solid line). The vertical dotted line highlights the position of the effective radius.}
\label{fig:profiles_unbiased_allin} 
\end{figure*}
In Fig. \ref{fig:profiles_unbiased_allin} we show the stacked void profiles before and after cleaning the void catalogues with our algorithm.
The stacked voids are all centred in the {\small VIDE} centres. We measure the cumulative number density of particles as a function of the distance from the centre, given in units of $r_\text{eff}$, for each of the selected voids, particle-per-particle from $0.3\ r_\text{eff}$ to $3.5\ r_\text{eff}$.
We divide this range of radii in 30 equally spaced bins and average the profiles.
This procedure is repeated before and after applying the cleaning algorithm.

From Fig. \ref{fig:profiles_unbiased_allin} it is possible to appreciate how differently we define the ridges of a void: the orange dashed profiles show the profiles of voids identified with {\small VIDE}, while the blue solid ones are for the same voids after cleaning and re-scaling.
The vertical dotted lines mark the radius of the void in both cases. We also show the background density level (solid horizontal black line) and the underdensity threshold of Eq.\eqref{eq:unbiased01} computed for the different redshifts (dashed horizonatal black line) as a reference.
It can be noticed that, while {\small VIDE} void limits reside close to the compensation wall and embed a density that approaches the background, our voids are, by construction, defined in much deeper regions of the density field.

These observations have to be kept in mind to understand how the algorithm we apply to clean the cosmic void catalogues affects the void structure.
Thanks to this modification we can be confident that the voids we are using to measure the size function are as close as possible to the objects for which we are able to construct the theoretical model. 
\begin{figure*}
 \centering
\includegraphics[width=0.95\textwidth]{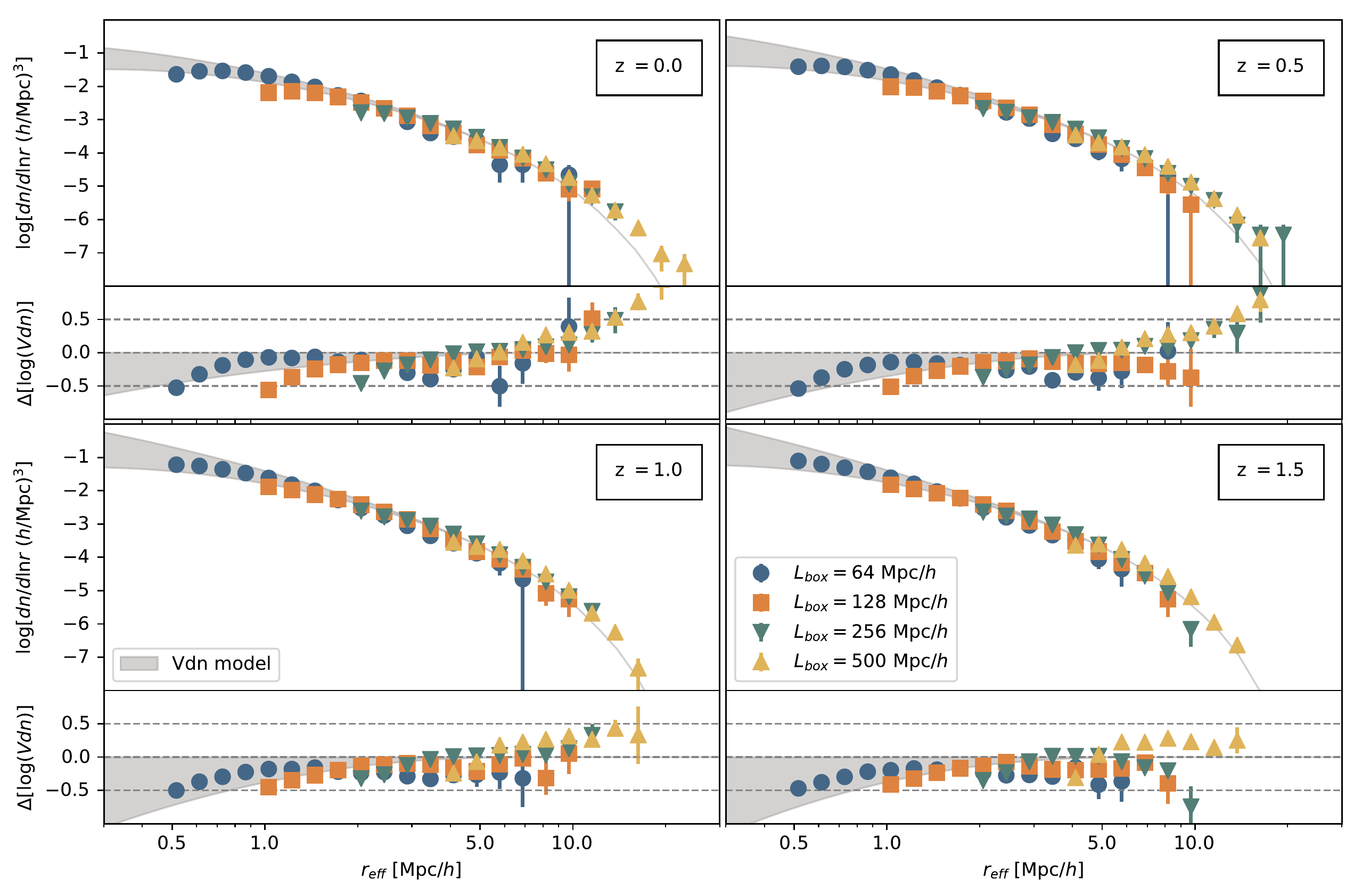}
\caption{{\em Upper part of each panel}: Void size function prediction (grey shaded region) at four different redshifts (different panels) compared to the distribution of cosmic voids after applying the cleaning method described in Sect. \ref{sec:finder} (different markers correspond to different simulations, the values of $L_{box}$ in legend corresponding to those of Table \ref{tab:simulations}). {\em Lower part of each panel}: Logarithmic differences between the measured distribution and the Vdn model prediction.}
\label{fig:sizefunc_unbiased_allin} 
\end{figure*}

Fig. \ref{fig:sizefunc_unbiased_allin} compares the measured and modeled void size functions, defined as the comoving number density of voids in logarithmic bins of effective radii, divided by the logarithmic extent of the bins. The four panels show the results at different redshifts. The different symbols in the upper part of each panel show the void size distributions measured in the simulation snapshots at different resolutions, while the grey shaded regions represent the model predictions.
The model depends on two thresholds, the one of the shell-crossing, which is fixed at the value $\delta_v^L = -2.71$, corresponding to our choice of threshold in the void catalogue cleaning, and the overdensity one, $\delta_{c}$, used to account for the {\em void-in-cloud} effect.
There is a persisting uncertainty on the definition of the latter, since it is not clear if this effect becomes relevant at turn-around (corresponding to a density contrast of $\delta_c \approx 1.06$) or when the overdensity collapses (at a density contrast $\delta_c \approx 1.69$).
Since in realistic occurrences we do not expect the spatial resolution of tracers to be high enough to inspect the radii range in which the void size function becomes sensitive to the {\em void-in-cloud} effect, we leave the overdensity threshold free to vary in the range $1.06 \le \delta_{c} \le 1.69$, thus the shaded region of Fig. \ref{fig:sizefunc_unbiased_allin}.

We cut the distribution measured in each of the simulations at twice the mean inter-particle separation (m.i.s.) of the given simulation box (see Table \ref{tab:simulations}).
To highlight the difference between the theoretical model predictions and the measured distributions, we show in the lower part of each panel the logarithmic difference between the simulation data and the model. 
Since voids are wide underdense regions, their measured properties and number density can be significantly affected by the spatial resolution of the sample. 
At scales comparable to the spatial resolution of the simulation, voids are not well represented by the mass tracers of the underlying density field. 
This causes a loss of power in the number counts, which can be noticed at the smallest scales in each panel of Fig.  \ref{fig:sizefunc_unbiased_allin}, in spite of the applied cut.
At large scales, instead, the measurements are limited by the simulation box extension. 
Since our cleaning algorithm does not consider the periodic boundary conditions of the N-body simulation yet, the largest peripheral objects of the box may exceed the box boundaries. As a consequence, we cannot trust voids close to the boundaries of the simulation box, and therefore we choose to reject them from our analysis.
Contrary to the limits at small radii, which cannot be overcome without increasing the resolution of the sample, this inaccuracy in the void counts at large radii could be faced with an upgrade of the void finding algorithm to account for the periodicity of the box.
Nonetheless, since our final goal is to provide a framework to exploit the distributions of voids observed in real catalogues, this issue is not addressed in this work and we simply cut away these objects and reduce the total volume of the box consistently.

Fig. \ref{fig:sizefunc_unbiased_allin} demonstrates that the Vdn model predictions are in overall good agreement with the results from N-body $\Lambda$CDM simulations, when cosmic voids are selected from unbiased matter tracers.
We notice that the model predictions are closer to the simulation measurements at higher redshifts, in the full range of investigated radii.
This might be due to the larger value of the underdensity threshold used for cleaning the catalogue: cosmic voids are less evolved at more distant cosmic epochs and have had less time to evacuate their interiors.
A higher threshold means a higher density of tracers, thus, a higher resolution.
Concerning the time evolution of the void size function, the density of the larger voids decreases with increasing redshifts. 
On the other hand, the smaller scales are less affected. 
The Vdn model well predicts this behaviour.

%%%%%%%%%%%%%%%%%%%%%%%%%%%%%%%%%%%%%%%%%%%%%%%%%%%%%%%%%%%%%%%%%%%%%%%%%%%%%%%
%%%%%%%%%%%%%%%%%%%%%%%%%%%%%%%%%%%%%%%%%%%%%%%%%%%%%%%%%%%%%%%%%%%%%%%%%%%%%%%

\subsection {Biased tracers}
\label {sec:biasedtracers}

Having established the reliability of the Vdn model in predicting the number counts of cosmic voids detected in a simulated  distribution of DM particles, we now extend our analysis to the case of biased tracers of the underlying DM density field. 
Our ultimate goal is to implement a model capable of predicting the void size distribution for observable tracers of the underlying density field, such as galaxies or clusters of galaxies.
In this paper we start focusing on the simplest case of biased tracers, namely DM haloes. 

It has been extensively demonstrated that voids in biased tracers of the underlying DM distribution are systematically larger than those predicted by the void size function models.
Specifically, the typical void sizes increase with the minimum mass of the tracers, that is with their effective bias \citep{Furlanetto2006}.

To infer cosmological constraints from the number count statistics of cosmic voids, the development of a reliable model independent of the tracers used to detect them is crucial.
In this section we make use of the set of halo catalogues obtained by means of a FoF algorithm applied to the N-body simulation snapshot at redshift $z = 0$ with volume $1\ (\text{Gpc}/h)^3$, whose properties are reported in Table \ref{tab:simulations}.

These catalogues are obtained by cutting the original FoF samples at $3$ different minimum masses: $M_{\text{min}} = \{2\cdot10^{12}\ M_\odot/h, 5\cdot10^{12}\ M_\odot/h, 10^{13}\ M_\odot/h\}$ (see Table \ref{tab:halocata}). These catalogues differ for the biasing factor with respect to the underlying DM distribution.

Our hypothesis is that the existing theoretical size function models fail in predicting the abundances of voids in the observed distributions of tracers because they implicitly assume that voids in biased tracers embed the same level of underdensity of the underlying DM distribution.
This is far from being true and can be shown by a simple analysis of the density profiles around the void centres.

We obtain the {\small VIDE} void catalogues from each of the DM halo catalogue reported in Table \ref{tab:halocata}, and clean them following the same procedure considered in Sect. \ref{sec:unbiasedtracers} for the unbiased DM distributions.
In particular, we keep the cleaning underdensity threshold parameter  unchanged, meaning that we set it to $\delta_v^{NL} = -0.795$ as we have done in Sec. \ref{sec:unbiasedtracers} for the unbiased catalogues at $z = 0$.
This choice is dictated by the need of selecting {\em sufficiently underdense} regions, to be distinguishable from noise, but at the same time {\em sufficiently dense} in tracers not to incur in resolution issues.
Nonetheless, we checked our final conclusions are robust with respect to the adopted threshold (Contarini et al. in preparation).

\begin{figure*}
    \centering
    \includegraphics[width=\textwidth]{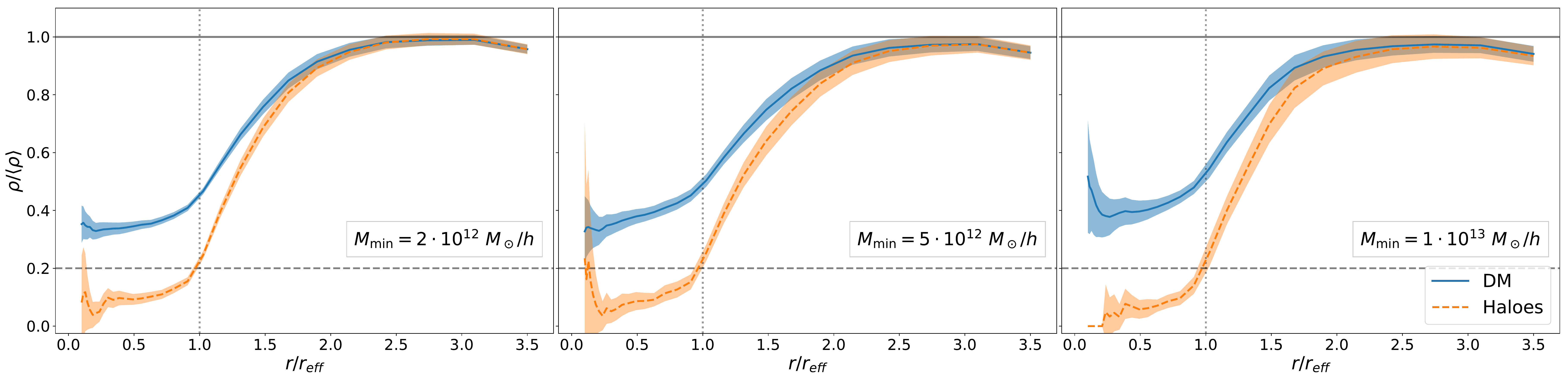}
    \caption{Stacked density profiles in different radii bins, for the three catalogues considered ( $M_\text{min} = 2\cdot10^{12}M_\odot/h$, $5\cdot10^{12}M_\odot/h$, $10^{13}M_\odot/h$, from left to right). Each panel shows the void stacked density profile. The dashed lines show the profiles measured in the DM halo distribution, while the solid lines represent the profile measured in the underlying DM distribution. The shaded bands around the mean profiles mark the $2\sigma$ confidence region. These density profiles have been measured using the cleaned {\small VIDE} void catalogues. The set of parameters used for the cleaning is the same one used for the unbiased DM voids in Sect. \ref{sec:unbiasedtracers}.}
    \label{fig:dens_prof_minnone_clean}
\end{figure*}

We then divide our catalogue of cosmic voids into a set of logarithmically equally spaced bins within the radii ranges reported in Table \ref{tab:halocata}, and compute the stacked density profile in each bin.
\begin{table*}
\centering
\footnotesize{\caption{Characteristic quantities for our set of halo catalogues. We report the minimum halo mass $M_\text{min}$, the total number of haloes $N_{H}$, the corresponding mean numerical density $\langle{n}\rangle$, the mean inter-particle separation (m.i.s.), the radii range used for profiling, the coefficients of the linear fit of Eq. \eqref{eq:linear_relation} ($b_\text{slope}$ and $c_\text{offset}$) with the value of the residuals (res.), and the ratio between the profile measured in the underlying unbiased DM distribution and in the DM halo distribution ($(\delta_\text{halo}/\delta_\text{DM})(r_\text{eff,h})$).} \label{tab:halocata}}
\begin{tabular}{lcrcrccccc}
\toprule	
$M_\text{min}$        & $N_{H}$  & $\langle{n}\rangle$   & m.i.s.         & radii range    & $b_\text{slope}$ & $c_\text{offset}$ & res.    & $(\delta_\text{halo}/\delta_\text{DM})(r_\text{eff,h})$ \\
$M_\odot/h$      & $10^{5}$ & $(\text{Mpc}/h)^{-3}$ & $\text{Mpc}/h$ & $\text{Mpc}/h$ &                  &                   &         &                                          \\
\midrule
$2\cdot10^{12}$  & $15.8$   & $1.58\cdot 10^{-3}$   & $8.58$         & $18\div40$     & $1.408$          & $0.022$           & $0.005$ & $1.406\pm0.001$                           \\
$5\cdot10^{12}$  &  $7.6$   & $7.62\cdot 10^{-4}$   & $10.95$        & $22\div45$     & $1.488$          & $0.013$           & $0.002$ & $1.522\pm0.001$                           \\
$10^{13}$  &  $3.9$   & $3.94\cdot 10^{-4}$   & $13.64$        & $27\div50$     & $1.657$          & $0.026$           & $0.008$ & $1.668\pm0.003$                           \\
\noalign{\smallskip}\hline
\bottomrule
\end{tabular}
\end{table*}
Figure \ref{fig:dens_prof_minnone_clean} compares the stacked density profiles of the voids selected in DM halo catalogues with three different mass cuts, measured in the halo and DM distributions. 
The profiles stacked in each bin of radii varies from a minimum of $5$ objects to a maximum of $300$ objects and depends on the total number of voids found by the algorithm in the considered range.
The number decreases going from smaller to larger radii.
In order to have a reasonable convergence of the mean profile, we only consider the stacked profiles for which we have at least 30 objects to average.
Figure \ref{fig:dens_prof_minnone_clean} shows also the background density and the shell-crossing underdensity level used to clean the catalogues which, by definition, crosses the profiles at $r = r_\text{eff}$.
The void density profiles measured from the distribution of DM haloes are significantly different with respect to those measured in the background matter density field, except at radii larger than about $2\, r_\text{eff}$, approximately where the background density level is recovered. 
Specifically, the density measured within voids in the DM halo distribution is way deeper than that measured in the underlying DM density field.

The clear discrepancy between the halo-void density profiles and the underlying DM density profiles is the reason why we cannot use the Vdn model directly to predict void abundances in biased tracers: theoretical models of the void size function are based on the evolution of underdensities in the DM density field and their shape severely depends on the void density contrast.
Ideally, we would like to recover the tracer density contrast which corresponds to the shell-crossing value.
However, since tracers are typically sparse, this is not practically viable: such a density contrast would be too low to be traced with enough statistics.
What we can do instead is to fix the density threshold in the tracer distribution and recover the corresponding density contrast in the underlying DM distribution.
Given the hypothesis that voids in tracers are centred in the same position of their DM counterparts, we search for the density contrast that voids identified in the tracer distribution reach in the underlying DM distribution.
By taking the ratio between the two profiles, as suggested by \cite{Pollina2017}, one can infer the relation between the density measured from biased tracers and the underlying unbiased DM density.

\begin{figure*}
    \centering
    \includegraphics[width=\textwidth]{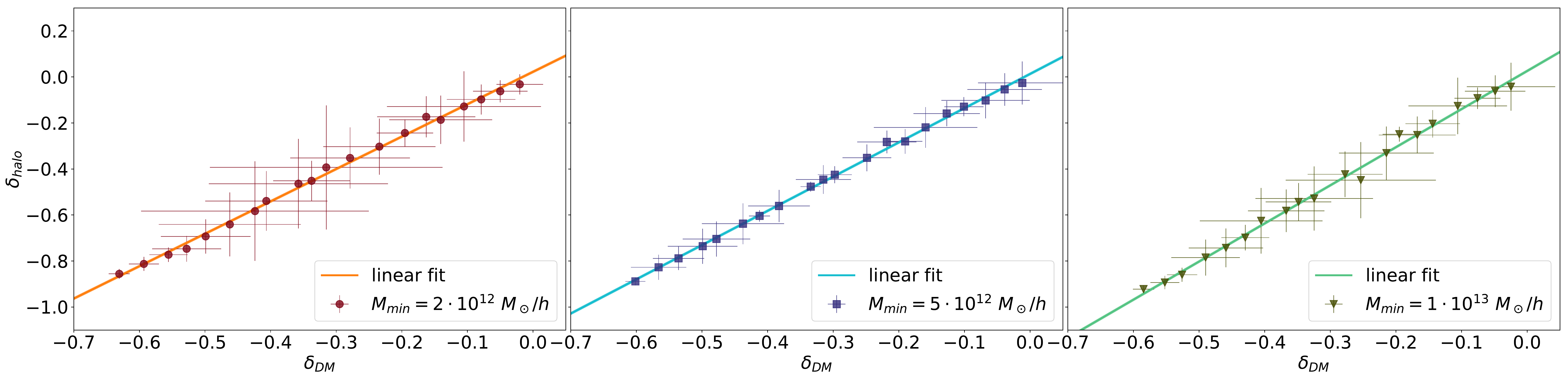}
    \caption{Density contrast in the unbiased tracer distribution, $\delta_\text{DM}$, versus density contrast in the biased tracer distribution, $\delta_\text{halo}$. Points are obtained gathering the measures of all the stacked profiles; the error bars represent a $1\sigma$ uncertainty on the mean values. The solid lines show the linear fit (the fitting parameters $b_\text{slope}$ and $c_\text{offset}$ are reported in Table \ref{tab:halocata}), while the narrow shaded regions show the uncertainty on the best-fit values given in terms of the residuals (also reported in Table \ref{tab:halocata}). Each panel shows the results for one of the halo catalogues of Table \ref{tab:halocata}, from left to right: $M_\text{min} = 2\cdot10^{12}M_\odot/h$, $5\cdot10^{12}M_\odot/h$, $10^{13}M_\odot/h$.}
    \label{fig:DM-HH_ratio}
\end{figure*}

Figure \ref{fig:DM-HH_ratio} shows the ratio between all the stacked profiles obtained for each catalogue, markers with error bars representing the uncertainty on the mean in the bin.
Given the high level of uncertainty in the inner regions of the profiles, due to the sparsity of tracers, we exclude all the values with $r \le 0.5\ r_\text{eff}$.
In agreement with the results by \cite{Pollina2017}, we find that the densities measured with the two different tracers, that is DM particles and haloes, are linearly related: 
\begin{equation}
    \label{eq:linear_relation}
    \delta_\text{halo} = b_\text{slope}\cdot\delta_\text{DM}+c_\text{offset} \,.
\end{equation}
The best-fit values of the two free parameters, $b_\text{slope}$ and $c_\text{offset}$, are reported in Table \ref{tab:halocata}.
The relation between the best-fit values of $b_\text{slope}$ and $c_\text{offset}$ to the large-scale effective bias of the tracer samples, $b_{eff}$, will be addressed in a forthcoming paper (Contarini et al. in preparation).

The existing theoretical models predict the comoving number density of underdense regions characterised by a given embedded density contrast in the DM density field.
Thus, to extend the models to the case of voids identified in the halo density field, we have to assess the density contrast in the underlying DM distribution.
This can be obtained by taking the ratio between the density profiles measured in the two different tracers.
Estimating it at exactly one effective radius allows us to associate the precise DM density contrast, required by the size function model for voids detected in distributions of biased tracers.

\begin{figure*}
    \centering
    \includegraphics[width=\textwidth]{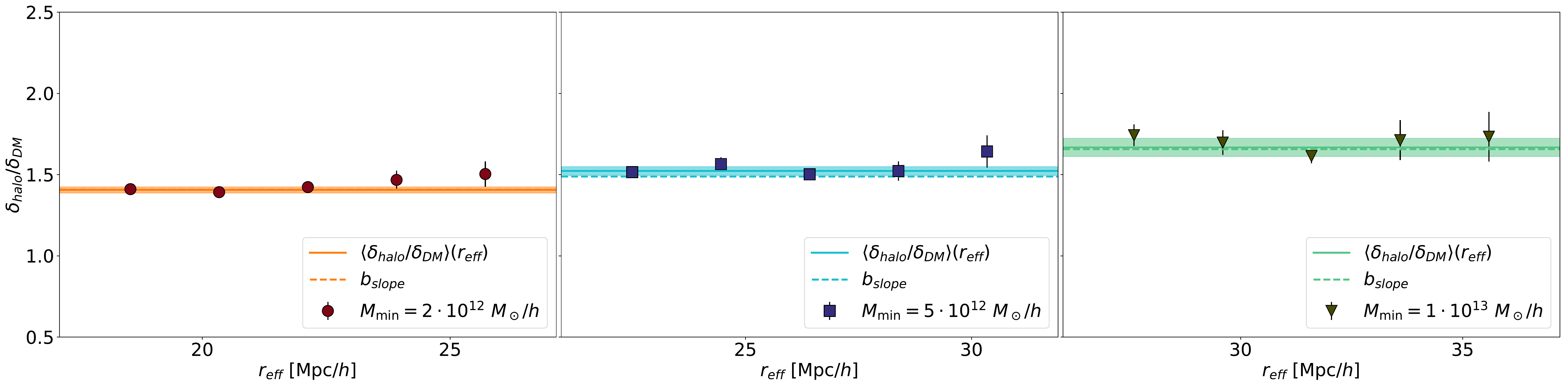}
    \caption{The ratio between $\delta_\text{halo}$ and $\delta_\text{DM}$ measured at $r = r_\text{eff}$ for voids with different sizes. Each point is obtained by measuring $\delta_\text{DM}(r_\text{eff})$ in different effective radius bins. The dashed lines are the values of $b_\text{slope}$ measured from the linear fit of Fig. \ref{fig:DM-HH_ratio}, while the solid lines are the average values, with the corresponding shaded regions delimiting $2 \sigma$ deviations from the mean. Each panel shows the results for one of the halo catalogues of Table \ref{tab:halocata}, from left to right: $M_\text{min} = 2\cdot10^{12}M_\odot/h$, $5\cdot10^{12}M_\odot/h$, $10^{13}M_\odot/h$.}
    \label{fig:deltaDMat1refHH}
\end{figure*}

Figure \ref{fig:deltaDMat1refHH} shows the ratio between the density contrast we have requested in the cleaning algorithm for halo-voids ($\delta_\text{halo}^{NL} = -0.795$) and the value of $\delta_\text{DM}^{NL}( r_\text{eff})$ measured from the three different catalogues.
Of all the stacked profiles we have obtained, we use for this measurement only those which counted at least $30$ objects in the radial bins to avoid issues with the convergence of the mean profile.
The mean value of this ratio is therefore the conversion factor between the underdensity thresholds used in the detection and cleaning of cosmic voids in our halo catalogues and the non-linear counterpart of the DM underdensity threshold required by the void size function theoretical model (the values of the mean and standard deviations are reported in Table \ref{tab:halocata}).
In the considered range of effective radii, the mean value of this ratio is consistent with what can be inferred from a linear fit, as also shown in Fig. \ref{fig:deltaDMat1refHH}.

Using the Vdn size function model, we can describe the distribution of voids in the DM distribution underlying the distribution of tracers.
If a cosmic void with radius $r_\text{eff}$ embeds a density contrast $\delta_{v, \text{DM}}^{\text{NL}}$ in the DM density field, then the same radius $r_\text{eff}$ will embed a density contrast $\delta_{v, \text{tr}}^{\text{NL}}$ in the tracer distribution, with $\delta_{v, \text{tr}}^{\text{NL}}$ and $\delta_{v, \text{DM}}^{\text{NL}}$ given by Eq. \eqref{eq:bias01}.
Therefore, if $\delta_{v, \text{tr}}^{\text{NL}}$ is the threshold we use when cleaning the original void catalogue (Sect. \ref{sec:finder}), then the resulting void size distribution has to be modelled by a size function with an underdensity threshold given by: 
\begin{equation}
\label{eq:biastr03}
\delta_{v, \text{DM}}^{\text{L}} = \mathcal{C}[1 - (1 + \delta_{v, \text{DM}}^{\text{NL}})^{-1/\mathcal{C}}]\text{,}
\end{equation}
where $\delta_{v, \text{DM}}^{\text{NL}}$ is the value shown in Fig. \ref{fig:deltaDMat1refHH}.

\begin{figure*}
    \centering
    \includegraphics[width=\textwidth]{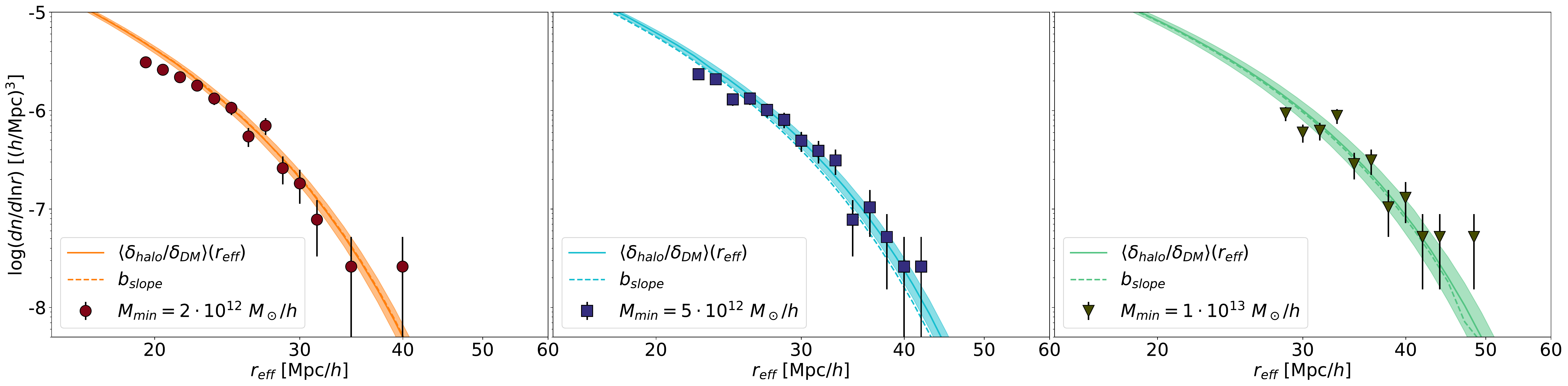}
    \caption{Size distribution of {\small VIDE} voids after the application of our cleaning prescription, measured in three different halo catalogues (markers). The shaded bands represent the $2\sigma$ confidence region around the mean values of the Vdn model modified to account for the different levels of DM underdensity enclosed by halo voids (solid lines). The dashed line shows the modified Vdn model in which instead we have used a threshold modified with the value of $b_\text{slope}$. Each panel shows the results for one of the halo catalogues of Table \ref{tab:halocata}, from left to right: $M_\text{min} = 2\cdot10^{12}M_\odot/h$, $5\cdot10^{12}M_\odot/h$, $10^{13}M_\odot/h$.}
    \label{fig:sizefunc_haloes}
\end{figure*}

Figure \ref{fig:sizefunc_haloes} shows the measured size functions of the voids detected in the three different DM halo catalogues, compared to the size function model described above.
The shaded band in each plot shows the $2\sigma$ confidence region around the average value of $\delta_{v, \text{DM}}^{\text{NL}}$, obtained from the analysis of the void density profiles.

We find an excellent agreement between the measured and predicted void size functions in all the considered catalogues (almost all the points are within the $2 \sigma$ confidence region), except at the smallest radii, where the spatial resolution of our catalogues does not allow us to have enough statistics of voids with $r_\text{eff} \lesssim 2.5\ \text{m.i.s.}$, thus causing a loss of power.
On the other hand, at large radii the size of the simulation box limits the possibility to obtain a reliable counting of large voids.
To investigate the impact of the offset parameter, $c_\text{offset}$, in Fig. \ref{fig:sizefunc_haloes} we also show the size function model obtained by considering only $b = b_\text{slope}$ as the correction parameter. This model systematically underestimates the measured void size function, though the mismatch is within $2 \sigma$, considering the estimated uncertainties.
This suggests that, even though it might not be the case when working with cosmic void density profiles, as stated in \cite{Pollina2017}, the value of the $c_\text{offset}$ coefficient of the relation cannot be ignored when computing the size function of cosmic voids.

The shell-crossing threshold provides a reasonable value to define a cosmic void. However, the size function model is not forced to be constructed using this threshold.
The model is in fact potentially capable of predicting the first-order statistics of density fluctuations whatever the threshold.
In particular, in this section we choose to use the shell-crossing threshold ($\delta_{v}^\text{NL}$) to clean the catalogues, selecting and rescaling the voids to the radius at which they reach this specific density contrast. Then we substitute in the theoretical size function (Eq. \eqref{eq:vsf01}) the threshold $\delta_v^L$ with the value that we measured from the stacked profiles ($\delta_\text{DM}$), converted to its linearly extrapolated counterpart by means of Eq. \eqref{eq:bias03}.
On the one hand, this is required for being representative of the underdensities embedded by the voids traced by biased distributions.
On the other hand, it also demonstrates that the use of the shell-crossing threshold is strictly required.
The detection of voids is prone to the nature of the tracer used for sampling the underlying DM distribution. Our work highlights that modelling their statistics cannot be done trivially without accounting for this nature.

Our overall conclusion is that the number count statistics of cosmic voids is completely determined by the cosmological model and by the relation between the density contrast of void tracers and DM {\em inside} the voids. This represents a key step towards the use of the void size function as a cosmological probe. 

%%%%%%%%%%%%%%%%%%%%%%%%%%%%%%%%%%%%%%%%%%%%%%%%%%%%%%%%%%%%%%%%%%%%%%%%%%%%%%%
%%%%%%%%%%%%%%%%%%%%%%%%%%%%%%%%%%%%%%%%%%%%%%%%%%%%%%%%%%%%%%%%%%%%%%%%%%%%%%%
%%%%%%%%%%%%%%%%%%%%%%%%%%%%%%%%%%%%%%%%%%%%%%%%%%%%%%%%%%%%%%%%%%%%%%%%%%%%%%%

\section{Summary}
\label{sec:conclusions}

The main goal of this work was to develop a theoretical model of the size function of cosmic voids detected from the distribution of biased tracers, extending the Vdn model, up to now validated for the DM distribution only.

The main steps of our analysis can be summarised as follows:

\begin{itemize}
\item We have developed a method to clean cosmic void catalogues, whose numerical implementation was presented in RM17. In particular, we have searched for the largest spherical regions embedding a mean density contrast of $0.2\; \overline{\rho}$, with $\overline{\rho}$ being the average density of the considered sample, ensuring that the identified regions do not overlap.
The condition on the mean density ensures that the selected underdensities have actually passed through shell-crossing, while the second condition guarantees the volume conservation.

\item We have validated the theoretical void size function using a set of DM catalogues extracted from N-body simulations with different characteristics in terms of box size and resolution.
We have verified that the predictions of the Vdn model are reliable when modelling the distribution of cosmic voids in the DM density field using the void samples cleaned with our method.
Indeed, we have found a good agreement between the model and the measurements for all the redshifts considered and on a large range of void radii.

\item To extend the theoretical size function model to the case of voids have extracted from biased mass tracers, we have built void catalogues from mock DM halo catalogues with different mass selections. Then we have computed the stacked density profiles, in a large range of radii, both in terms of density of haloes and of the underlying DM distribution. 
\item We have found a linear relation between the halo-profiles and the DM-profiles inside cosmic voids, in agreement with \citet{Pollina2017}.
\item By measuring the ratio between the density contrast of stacked profiles in haloes and DM at $r=r_\text{eff}$, we have found the value of the multiplicative constant $b$, to be used to model the cosmic void size function in biased distributions (Eq. \eqref{eq:bias01}), as the ratio between the density profiles measured in the tracers distribution and the density profiles measured in the underlying DM distribution, $(\delta_{halo}/\delta_{DM})(r_{eff})$, at one effective radius $r_{eff}$.
\item Finally, we have compared the proposed size function model to the ones measured from DM halo catalogues with different mass cuts, finding an excellent agreement.
\end{itemize}

In a forthcoming project we will further extend the proposed size function model, parameterising it as a function of the large-scale effective bias of the tracers used to identify the voids. Furthermore, we will provide forecasts on the cosmological constraints than can be inferred from number count statistics of cosmic voids (Contarini et al. in preparation).

\section*{Acknowledgements}
TR is thankful to Alfonso Veropalumbo for useful discussions.
TR is also grateful for the support of his supervisors, Andrea Lapi and Matteo Viel.
FM and LM  acknowledge the grants ASI n.I/023/12/0, ASI-INAF n.
2018-23-HH.0 and PRIN MIUR 2015 ``Cosmology and Fundamental Physics: illuminating the Dark Universe with Euclid''. 
MB acknowledges support from the Italian Ministry for Education, University and Research (MIUR) through the SIR individual grant SIMCODE, project number RBSI14P4IH. The N-body simulations described in this work have been performed on the Hydra supercomputer at RZG and on the Marconi supercomputer at Cineca thanks to the PRACE allocation 2016153604 (P.I. M. Baldi).

%%%%%%%%%%%%%%%%%%%%%%%%%%%%%%%%%%%%%%%%%%%%%%%%%%


\bibliographystyle{mnras}
\bibliography{main}

%%%%%%%%%%%%%%%%%%%%%%%%%%%%%%%%%%%%%%%%%%%%%%%%%%

%%%%%%%%%%%%%%%%%%%%%%%%%%%%%%%%%%%%%%%%%%%%%%%%%%

% Don't change these lines
\bsp	% typesetting comment
\label{lastpage}
\end{document}